\documentclass{aa}
\usepackage{psfig}

\begin{document}
\title{A 1.2~mm MAMBO/IRAM-30m Study of Dust Emission from optically
luminous $z \approx 2$ Quasars}

\author{Alain Omont\inst{1}
        \and Alexandre Beelen\inst{2}
        \and Frank Bertoldi\inst{3}
        \and Pierre Cox\inst{2}
        \and Chris L. Carilli\inst{4}
        \and \\ Robert S. Priddey\inst{5}
        \and Richard G. McMahon\inst{6}
        \and Kate G. Isaak\inst{7}
        }

\offprints{A. Omont, omont@iap.fr}

\institute{
     Institut d'Astrophysique de Paris, CNRS, 98bis boulevard Arago, 
     F-75014 Paris, France
\and Institut d'Astrophysique Spatiale,  Universit\'e de Paris XI, 
     F-91405 Orsay, France
\and Max-Planck-Institut f\"ur Radioastronomie, Auf dem H\"ugel 69, 
     D-53121 Bonn, Germany
\and National Radio Astronomy Observatory, P.O. Box O, Socorro, 
     NM 87801, USA
\and Blackett Laboratory, Imperial College of Science, Technology 
     \& Medicine, Prince Consort Road, London SW7 2BW, UK
\and Institute of Astronomy, Madingley Road, Cambridge CB3 0HA, UK
\and Cavendish Laboratory, Madingley Road, Cambridge CB3 0HE, UK
}

\date{Received date June 26, 2002/ Accepted date October 2002}

\titlerunning{A 1.2~mm MAMBO survey of $z \approx 2$ quasars}
\authorrunning{A. Omont et al.}

\abstract{We report 250~GHz (1.2~mm) observations of redshift $1.8 < z
  < 2.8$ optically luminous ($\rm M_B < -27.0$), radio quiet quasars using
  the Max-Planck Millimeter Bolometer (MAMBO) array at the IRAM
  30-metre telescope. Nine quasars were detected and for 26 quasars
  3$\,\sigma$ flux density limits in the range 1.8 to 4~mJy were
  obtained.  Adopting a typical dust temperature of 45~K, the
  millimeter emission implies far-infrared luminosities of order $\rm
  10^{13} \, L_{\rm \odot}$ and dust masses of $\sim 10^8 \, \rm
  M_{\rm \odot}$.  Applying a statistical survival analysis to our total
  sample of 43 detected and 95 undetected quasars at $z \approx 2$ and
  $z \ga  4$, we find that there is no apparent difference in the
  far-infrared (FIR) luminosities, hence the star formation rates, of QSOs
  at $z \approx 2$ and at $3.6< z <5$.  This differs from radio
  galaxies, for which the FIR luminosity was found to increase with
  redshift (Archibald et al.\ 2001). We furthermore find that there is
  no strong correlation between the far-infrared and optical
  luminosities, confirming previous results obtained on smaller
  samples.   }

\maketitle

\sloppy

\section{Introduction}
\label{sec:introduction}

The relation between the growth of the central black hole and the
formation of the bulge stars is a key issue for the formation and
evolution of galaxies. Probing starburst activity in the host galaxies
of high redshift quasars has therefore become a key area of
observational cosmology.  Since the formation of massive stars often
occurs in heavily obscured regions, star formation is best traced
through the far-infrared (FIR) dust emission (e.g., Sanders \& Mirabel
1996), the peak of which is red-shifted into the (sub)millimeter
atmospheric windows for sources at redshifts $z > 1$.  Deep
(sub)millimeter blank field surveys using SCUBA (e.g., Scott et al.
2002) and MAMBO (Bertoldi et al. 2000a,b) have now detected over one
hundred sources. For most of these objects optical obscuration
prevents a spectroscopic redshift determination and any detailed study
of the relation between star formation and Active Galactic Nuclei
(AGN). Therefore, such studies must currently rely on pointed
(sub)millimeter observations of radio galaxies and of optically
selected quasars.

Observations at 1.2 mm and 0.85 mm using MAMBO and SCUBA,
respectively, have lead to the detection of more than 60 high-redshift
quasars (Omont et al.  1996 and references therein for earlier work;
Omont et al.  2001; Carilli et al.  2001a; Isaak et al. 2002; and present paper; Priddey et al. 2002), radio
galaxies (Archibald et al. 2001), and X-ray selected AGN (Barger et
al.  2001; Page et al.  2001). Whether most of the dust FIR emission
is due to heating by the UV/X radiation of the AGN or by massive stars
remains unclear.  Only for a few cases there are searches for CO line emission deep enough for achieving detection and providing good evidence that a substantial
fraction of the FIR emission must be caused by star formation
(Guilloteau et al. 1999; Cox et al.  2002a,b; Carilli et al. 2002).

Up to now most of the studies of the (sub)millimeter emission from
high redhift QSOs have focused on sources at redshifts $z \ga  4$, for
which 40 detections were reported (Omont et al. 1996, Omont et al.
2001, Carilli et al. 2001a, Isaak et al. 2002, and references therein). In order to study
the quasar population at the peak of their space density, which is at
a later epoch than that probed in previous studies, we extended the
millimeter observations of optically luminous quasars to $z \approx
2$. This paper reports on 1.2~mm continuum observations of a sample of
redshift 1.8 to 2.8 QSOs with optical luminosities in excess of
$10^{14} \, L_\odot$ ($\rm M_B < -27.0$).

We adopt a $\Lambda$-cosmology with $\rm H_0=65\rm~km~s^{-1}~Mpc^{-1}$,
$\rm \Omega_\Lambda=0.7$ and $\rm \Omega_m=0.3$. However, for the definition
of the rest-frame absolute B-band magnitudes, $\rm M_B$, we have used the
standard Einstein-de Sitter cosmology with $\rm H_0 = 50$~$\rm km\,s^{-1}$
Mpc$^{-1}$ and $\rm q_0$=0.5, in order to ease the comparisons with
standard quasar catalogues and luminosity functions.  Throughout this
paper, the far-infrared luminosity ($L_{\rm FIR}$) is defined as the
luminosity of a modified black-body with a dust temperature and
emissivity index as given in Sect.~\ref{sec:results:properties}.

\section{Source Selection and Observations}
\label{sec:observations}

Thirty-five optically luminous ($\rm M_B < -27.0$) radio-quiet QSOs at
redshifts $1.8 < z < 2.8$ were observed from 60 sources we had
selected from the catalogue of V\'eron-Cetty \& V\'eron (2000). As the
result of our random selection among the sources listed by
V\'eron-Cetty \& V\'eron, the observed sources derive from a broad
variety of QSO surveys. However, about half of them were originally
identified in the Hamburg Quasar Survey (Engels et al. 1998; Hagen;
Engels \& Reimers 1999, http://www.hs.uni-hamburg.de). The observed
sources have a distribution in rest-frame absolute B-band magnitudes,
$\rm M_B$, between $-27.0$ and $-29.5$, which is comparable to the $\rm M_B$
distribution of $z \ga  4$ PSS quasars studied by Omont et al. (2001)
-- see Sect.~\ref{sec:results:properties}. For the present sample at
z~$\approx$~2, the values of $\rm M_B$ were determined including J and H
magnitudes obtained from the 2MASS near-infrared survey - see Priddey
et al. (2002). However, the $\rm M_B$ values are more uncertain
for the $z \ga  $4 sample because 2MASS data are lacking for most
sources.  The selected QSOs are at high declination ($> 15^\circ$ and
mostly $> 35^\circ$) to favor observations done at low air mass. The
list of the observed sources is given in Tables 1 and
2.

The observations were made during the winter of 2000-2001 using the
37-channel {\it Max-Planck Millimeter Bolometer} (MAMBO; Kreysa et al.
1999a, 1999b) array at the 30-meter IRAM telescope on Pico Veleta
(Spain).  The sources were observed with the array's central channel,
using the standard on-off mode with the telescope secondary chopping
in azimuth by 50$^{\prime\prime}$ at a rate of 2$\,$Hz.  The target
was positioned on the central bolometer of the array, and after 10
seconds of integration, the telescope was nodded so that the previous
{\it off} beam became the {\it on} beam. A typical scan lasts for 12
or 16 such 10 seconds sub-scans.  The pointing was checked frequently
on nearby continuum sources, and was found to be stable within $\sim
2^{\prime\prime}$.  The sky opacity was measured regularly, with
zenith opacities at 1.2~mm varying between 0.08 and 0.4. Gain
calibration was performed using observations of Mars.  We adopted a
calibration factor of 12500 counts per Jansky, which we estimate to be
reliable to within 20\%. The point-source sensitivity of MAMBO during
the observations was $\sim 30 \, \rm mJy \, s^{1/2}$.

The data were analyzed using the MOPSI software (Zylka 1998).
Correlated sky-noise was subtracted from each channel; it was computed
for each channel as a weighted mean of the signals from the eight
best-correlating surrounding channels. Skynoise subtraction reduces
the noise in the signals by typical factors of 2 to 3. For each
quasar, the total on-target plus off-target observing time was always
greater than 650~sec.  The mean r.m.s noise of the coadded signals is
$\rm \approx 0.8 \, mJy$.  As for Omont et al. (2001), we defined a
quality factor for the detections (A for good and B for poor) based on
the consistency of the flux densities measured on different dates, on
the stability of the pointing, on calibration uncertainties and
general observing conditions (see
Table 1).
        
                %% Table 1
        
% Table 1 listing the detected quasars
%% redone on oct 9 2001
\begin{table*}[tbhp]
\caption{Quasars detected at 1.2~mm. 
         \label{detections}}
 \begin{center}
\begin{tabular}{lccccrrcc}
\hline
Source & z & $M_B$ & R.A. & Dec. & Flux Density  & time & Quality  &  1.4~GHz \\ 
  & & & \multicolumn{2}{c}{(J2000.0)} & $\mbox{[mJy],}\pm 1 \sigma$ & [sec] &  & $\mbox{[mJy],}\pm 1 \sigma$\\ 
\hline

KUV 08086+4037                  & $ 1.78 $ & $ -27.0 $ & 08 12 00.5 &  +40 28 14.0 & $  4.3 \pm 0.8 $ & $ 1336 $ & A & --\\

[VV96] J093750.9+730206         & $ 2.52 $ & $ -28.5 $ & 09 37 48.7 &  +73 01 58.0 & $  3.8 \pm 0.9 $ & $  665 $ & B & *\\

HS 1002+4400                    & $ 2.08 $ & $ -28.3 $ & 10 05 17.5 &  +43 46 09.0 & $  4.2 \pm 0.8 $ & $ 1654 $ & A & --\\

HS 1049+4033                    & $ 2.15 $ & $ -28.2 $ & 10 51 58.6 &  +40 17 36.0 & $  3.2 \pm 0.7 $ & $ 1818 $ & A & --\\

[VV96] J110610.8+640008         & $ 2.19 $ & $ -29.3 $ & 11 06 10.8 &  +64 00 08.0 & $  3.9 \pm 1.1 $ & $ 2228 $ & A & *\\

[VV96] J140955.5+562827         & $ 2.56 $ & $ -28.4 $ & 14 09 55.5 &  +56 28 27.0 & $ 10.7 \pm 0.6 $ & $ 1832 $ & A & --\\

[VV96] J154359.3+535903         & $ 2.37 $ & $ -28.3 $ & 15 43 59.3 &  +53 59 03.0 & $  3.8 \pm 1.1 $ & $  992 $ & B & --\\

HS 1611+4719                    & $ 2.35 $ & $ -27.7 $ & 16 12 39.9 &  +47 11 58.0 & $  4.6 \pm 0.7 $ & $ 3319 $ & A & --\\

[VV2000] J164914.9+530316       & $ 2.26 $ & $ -28.2 $ & 16 49 14.9 &  +53 03 16.0 & $  4.6 \pm 0.8 $ & $ 1655 $ & A & $0.70\pm0.09$ \\

\hline
\end{tabular}
\end{center}
NOTE -- The radio flux densities at 1.4~GHz are extracted from the VLA FIRST survey.
The symbol * indicates that \\
there is no data available, and the minus symbol indicates
a $6 \, \sigma$  upper limit to the flux density of 1~mJy.    
\end{table*}

        \label{Table1}

                %% Figure 1
        \begin{figure}[]     
        \centerline{\psfig{figure={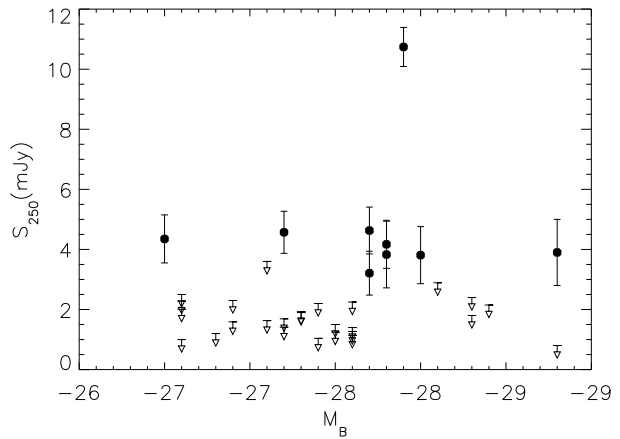},width=9cm}}
        \caption{Observed 1.2~mm flux density versus $\rm M_B$, the
          optical absolute magnitude in the rest-frame B band, of the
          $z\approx 2$ QSOs. The sources detected at 1.2~mm
          (Table1) are shown as filled symbols. The open
          symbols are upper limits for sources not detected at 1.2~mm
          (Table2): for sources with negative signal, a
          $1\,\sigma$ upper limit is plotted; for sources with
          positive signal, $\rm signal + 1\sigma$ is plotted as an
          upper limit.}
        \label{figure1}
        \end{figure}

\section{Results}

\subsection{General properties}
\label{sec:results:properties}

Thirty-five $z \approx 2$ quasars were observed at 1.2~mm.  Nine
quasars were detected at levels $\ge 3 \,\sigma$, of which seven have
signals $\ga  4 \,\sigma$, where $\sigma \approx 0.8 \, \rm mJy$. The
1.2~mm flux densities of these quasars are listed in
Table 1. Twenty-six sources were not detected with $3 \,
\sigma$ flux density upper limits in the range 1.8--4~mJy
(Table 2). The weighted mean signal of the non-detections
is $\rm 0.71 \pm 0.16 \, mJy$.

Fig.~\ref{figure1} displays the observed 1.2~mm flux density $S_{\rm 250}$
versus $\rm M_B$, the optical absolute magnitude in the B rest-frame band,
which represents the bolometric luminosity, $L_{\rm bol}$. Leaving aside
J1409, which has an exceptionally high 1.2~mm flux density of $10.7
\pm 0.6 \, \rm mJy$, the other eight detected quasars have comparable
flux densities of 3--4~mJy.  There is no obvious correlation between
$\rm M_B$ and $S_{\rm 250}$, a point which will be discussed in
Sect.~\ref{sec:statistics}.

%% Table 2
	%% Table 2 - Source whihc were not detected
\begin{table*}[tbhp]
\caption{Quasars with upper limits at 1.2~mm. 
         \label{nondetections}} 
\begin{center}
\begin{tabular}{lccccrrc}
\hline \\[-0.2cm]
Source & z & $M_B$ & R.A. & Dec. & Flux Density & time & 1.4~GHz \\ 
  & & & \multicolumn{2}{c}{(J2000.0)} & $\mbox{[mJy],}\pm 1 \sigma$ & [sec] & $\mbox{[mJy],}\pm 1 \sigma$\\ 
\hline \\[-0.2cm]

HS 0749+4259 		& $ 1.90 $ & $  -28.9 $ & 07 50 54.7 & +42 52 19.0 & $ 1.3 \pm 0.9 $ & $ 1652 $ & --\\

HS 0800+3031$^\dagger$	& $ 2.02 $ & $  -28.6 $ & 08 03 42.1 & +30 22 54.0 & $ 1.8 \pm 1.1 $ & $  824 $ & --\\

HS 0808+1218$^\dagger$	& $ 2.26 $ & $  -27.8 $ & 08 10 57.0 & +12 09 14.0 & $ 1.1 \pm 0.9 $ & $ 1583 $ & --\\

HS 0830+1833$^\dagger$	& $ 2.27 $ & $  -28.1 $ & 08 32 55.7 & +18 23 01.0 & $ 1.3 \pm 0.9 $ & $ 1414 $ & --\\

HS 0834+1509$^\dagger$	& $ 2.51 $ & $  -28.1 $ & 08 37 12.6 & +14 59 17.0 & $ 0.6 \pm 0.6 $ & $ 2391 $ & --\\

SBSS 0910+586		& $ 1.95 $ & $  -27.1 $ & 09 14 25.8 & +58 25 19.0 & $ 0.4 \pm 0.6 $ & $  839 $ & * \\

[VV96] J092230.1+710937	& $ 2.43 $ & $  -27.1 $ & 09 22 30.1 & +71 09 37.0 & $ 1.5 \pm 0.8 $ & $ 1666 $ & * \\

HS 0932+2410		& $ 2.30 $ & $  -27.1 $ & 09 35 34.0 & +23 57 20.0 & $ 1.4 \pm 0.6 $ & $ 2074 $ & --\\

[VV96] J093935.1+361001	& $ 2.03 $ & $  -27.1 $ & 09 39 35.1 & +36 40 01.0 & $ 1.5 \pm 1.0 $ & $ 1498 $ & --\\

[VV96] J095845.5+470324	& $ 2.48 $ & $  -27.7 $ & 09 58 45.5 & +47 03 24.0 & $ 0.7 \pm 0.7 $ & $ 1647 $ & --\\

HS 1110+3813		& $ 2.29 $ & $  -28.1 $ & 11 12 51.0 & +37 57 15.0 & $ 0.0 \pm 1.3 $ & $  833 $ & --\\

KUV 11467+3725		& $ 2.22 $ & $  -27.8 $ & 11 49 20.2 & +37 08 29.0 & $ 1.0 \pm 0.9 $ & $ 1667 $ & --\\

[VV96] J121010.2+393936	& $ 2.40 $ & $  -27.7 $ & 12 10 10.2 & +39 39 36.0 & $ 0.9 \pm 0.8 $ & $ 1492 $ & --\\

[VV96] J121303.1+171422$^\dagger$	& $ 2.54 $ & $  -28.0 $ & 12 13 03.1 & +17 14 22.0 & $ -1.9 \pm 1.5 $ & $  656 $ & $1.97\pm0.16$  \\

[BCF91] 524		& $ 2.85 $ & $  -28.0 $ & 13 04 12.0 & +29 53 49.0 & $ 0.2 \pm 1.0 $ & $  329 $ &  --\\

[BBE90] 130623+283002	& $ 2.21 $ & $  -27.9 $ & 13 09 17.2 & +28 14 04.0 & $ -1.6 \pm 1.0 $ & $ 1648 $ & --\\

[VV96] J140148.4+543859	& $ 2.37 $ & $  -27.3 $ & 14 01 48.4 & +54 38 59.0 & $ 0.3 \pm  0.9 $ & $  842 $ & --\\

SBSS 1417+596		& $ 2.31 $ & $  -27.6 $ & 14 19 06.4 & +59 23 12.0 & $  -3.1 \pm  1.6 $ & $  831 $ & * \\

[VV96] J160637.6+173516	& $ 2.32 $ & $  -27.4 $ & 16 06 37.6 & +17 35 16.0 & $ 1.6 \pm  0.7 $ & $ 2176 $ & --\\

HS 1616+3708		& $ 2.49 $ & $  -27.4 $ & 16 18 15.5 & +37 01 03.0 & $ 1.0 \pm  0.6 $ & $ 2831 $ & --\\

[VV86] J162645.7+642654	& $ 2.32 $ & $  -28.8 $ & 16 26 45.7 & +64 26 54.0 & $ 0.7 \pm  1.1 $ & $ 3470 $ & *\\

HS 1707+4602		& $ 2.29 $ & $  -27.6 $ & 17 09 04.9 & +45 59 08.0 & $ 2.1 \pm  1.5 $ & $ 2487 $ & --\\

[VV96] J171635.4+532815	& $ 1.94 $ & $  -28.8 $ & 17 16 35.4 & +53 28 15.0 & $ 1.3 \pm  1.1 $ & $ 2469 $ & $1.90\pm0.16$ \\

HS 1754+3818$^\dagger$	& $ 2.16 $ & $  -27.9 $ & 17 56 39.6 & +38 17 52.0 & $ 1.1 \pm  1.1 $ & $ 3282 $ & *\\

[VV96] J183825.0+510558	& $ 1.98 $ & $  -29.3 $ & 18 37 25.3 & +51 05 59.0 & $-1.0 \pm  0.8 $ & $ 2452 $ & * \\

HS 2134+1531$^\dagger$	& $ 2.13 $ & $  -28.1 $ & 21 36 23.7 & +15 45 08.0 & $-2.7 \pm  1.4 $ & $ 2494 $ & *\\

\hline \\[-0.2cm]
\end{tabular}
\end{center}
NOTE -- The radio flux densities at 1.4~GHz are extracted from the VLA FIRST survey.
Symbols as for Table~1. \\ 
$^\dagger$ Quasars observed (and not 
detected) at 850~$\rm \mu m$ by Priddey et al. (in preparation).
\end{table*}

\label{Table2}
        
With a single measurement at 250~GHz (at $z \sim 2.2$, this
corresponds to an emitted frequency of $\rm \sim 800 \, GHz$ or a
wavelength of $\rm 375 \, \mu m$), it is not possible to distinguish
whether the observed millimeter emission is thermal dust or
synchrotron emission.  The VLA 1.4~GHz FIRST survey (Becker, White, \&
Helfand 1995) provides a first estimate on the radio to millimeter
slope. No radio source within 30$^{\prime\prime}$ of six of the $z
\approx 2$ detected quasars is found to a $6 \, \sigma$ limiting flux
density of $\rm \approx 1 \, mJy$. Two mm-detected QSOs are not
covered by the FIRST survey, and J1649 coincides with a radio source
with a 1.4~GHz flux density of $\rm 0.70\pm 0.09 \, mJy$
(Table 1).  For the quasars not detected at 1.2~mm
(Table 2), two sources have a radio counterpart with
1.4~GHz flux densities of $\rm \approx 2 \, mJy$.  In all cases, the
1.4~GHz flux densities are too small to consider the quasars as radio
loud. For the detected source J1649, the millimeter-radio spectral
index is $\approx 0.37$, well above the extrapolation of a synchrotron
spectrum and close to the value expected at $z \approx 2$ for
starbursts (e.g., Carilli \& Yun 1999). Its case is similar to the
QSOs PSS~J1048+4407 and PSS~J1057+4555 detected at 1.2~mm and 1.4~GHz
with comparable flux densities (Omont et al. 2001; Carilli et al.
2001b). We shall assume in the following that the millimeter flux
density of the $z\approx 2$ quasars detected at 1.2~mm is thermal in
nature, as supported by their very recent deep VLA 1.4~GHz observations
(Petric et al. in preparation).

%%Further confirmation of this assumption will need deeper
%%measurements of the radio emission and submillimeter observations.

For consistency with our earlier treatments (Omont et al. 2001), we
adopt a dust emissivity index of $\beta = 1.5$, and a dust
temperatures of 45~K\footnote{Adopting different values for the dust
  temperature and $\beta$, e.g., 40~K and 2.0 (Priddey \& McMahon
  2001), will not change the main conclusions of this paper}.  In the
redshift range $2 < z < 3$, the FIR luminosity can then be expressed
as a function of the 1.2~mm (250~GHz) flux density, $S_{\rm 250}$, as
   \begin{equation}
   L_{\rm FIR} \sim  4.7 \times 10^{12} ~(S_{\rm 250}/ {\rm mJy} ) ~~ \rm L_\odot ~.
   \end{equation}
\noindent
The corresponding proportionality factor for the $z \ga  4$ quasars is
$3.5 \times 10^{12}$.  The derived FIR luminosities for the $z\approx
2$ quasars detected at 1.2~mm are $\approx 10^{13} \rm \, L_\odot$,
with corresponding dust masses of a few $10^8 \times
(7.5/\kappa_{\rm d230}) \, \rm M_\odot$, where $\kappa_{\rm d230}$ is the dust
absorption coefficient at $\rm 230 \, \mu m$ in units of $\rm cm^2 \,
g^{-1}$.

Like for the QSOs detected at $z \ga  4$, for the newly detected objects
$L_{\rm FIR}$ is about one tenth of the optical luminosity $L_{\rm opt}$. If a
substantial fraction of $L_{\rm FIR}$ arises from young stars, such
luminosities imply star formation rates approaching $10^3 \, \rm
M_\odot \, yr^{-1}$, comparable to the values derived for $z \ga   4$
quasars (Omont et al.  2001; Carilli et al. 2001a).

\subsection{Comments on Individual Sources}
\label{sec:results:individual}
        
\begin{itemize}

\item{VV96 J140955.5+562827 (SBS 1408+567)}
  
  This quasar has an exceptionnally strong 1.2~mm flux density which
  puts it among the seven known high $z$ sources with S$_{\rm 250} \, \ga 
  \, 10 \, \rm mJy$. Four  of the latter are known to be strongly
  lensed, while there is presently no indication of a strong
  amplification for SBS~1408+567. The optical spectrum of SBS~1408+567
  shows spectacular detached broad absorption lines (BAL) with double
  troughs (Korista et al. 1993). Deriving the redshift from the
  unpublished Keck HIRES spectrum of Barlow \& Junkkarinen (1994) -
  and Junkkarinen (private communication), we recently detected CO in
  this QSO using the IRAM Plateau de Bure interferometer (Beelen et
  al.  in preparation).

\smallskip

\item{VV96 J154359.3+535903 (SBS 1542+541)}
  
  This source was discovered in the Second Byurakan Survey (Stepanyan
  et al. 1992) and has many interesting properties: BAL with a very
  high ionization degree (Telfer et al. 1998), associated absorption
  system and damped Ly$\alpha$ (DLA) absorption system, and a strong
  X-ray absorption (Green et al. 2001).  
  \smallskip

\item{VV96 J110610.8+640008 (HS 1103+6416)}
  
  It is the brightest optical source of the sample. Its spectrum was
  analysed by Koehler et al. (1999), using a high resolution
  HIRES/Keck optical spectrum and a UV HST spectrum. The most striking
  feature is a strong, complex absorption system at $z = 1.9$.
  \smallskip

\item{KUV 08086+4037}
  
  The only reference in SIMBAD is the discovery paper (Darling \&
  Wegner 1996), quoting broad emission lines. 
  \smallskip

\item{VV2000 J093748.7+730158}
  
  It displays a rich absorption system with a complex DLA at $z =
  1.478$ and a Lyman limit system near $z = 2.36$ (Rao \& Turnshek
  2000).
  \smallskip

\item{HS 1002+4400, HS 1049+4033 and HS 1611+4719}
  
  For these three sources, the only references in SIMBAD are the
  report of their discovery in the Hamburg survey (Hagen, Engels \&
  Reimers 1999). Their low--resolution spectra (available at
  http://www.hs.uni-hamburg.de) display rather broad emission lines
  without very peculiar features.
  \smallskip

\item{VV2000 J164914.9+530316}
  
  There is not a single explicite reference in SIMBAD except the
  V\'eron-Cetty \& V\'eron catalogue. As discussed in
  Sect.~\ref{sec:results:properties}, this QSO is weakly radio loud.

\end{itemize}

\section{Statistical Study}
\label{sec:statistics}

In order to compare the results of the 1.2~mm studies of the $z
\approx 2$ and $z \ga  4$ quasars and to search for possible correlations
between the FIR luminosity (derived from the 1.2~mm flux density) and
the optical luminosity (derived from the absolute B-Band magnitude
$\rm M_B$), we made a statistical study of the present data set and the $z
\ga  4$ QSO surveys by Omont et al. (2001) and Carilli et al. (2001a)
which are based on the PSS and SDSS QSO samples, respectively, and are
hereafter referred to as the PSS and SDSS samples.
The statistical study is based on the survival analysis method which
is the most appropriate to handle the {\em censored data} (upper or
lower limits) in astronomical surveys (Feigelson \& Nelson 1985;
Isobe et al. 1986, Isobe \& Feigelson 1986). In these methods the values of the
upper (or lower) limits are formally taken into account for the
statistical analysis. However, the algorithms assume that the limiting
value is precisely measured, while in astronomy the limiting value is
rather a specific likelihood upper limit (e.g., Gleser 1992). In this
paper, we adopt a $3\, \sigma$ (99\%) upper limit for the survival
analysis whereas in the figures of this paper the upper limits are
shown at the 68\% level confidence level (see caption of
Fig.~\ref{figure1}).

\subsection{Properties of the $z \approx 2$ and $z \geq  4$ QSOs}
\label{sec:statistics:properties}

\begin{figure}[]
  \centerline{\psfig{figure={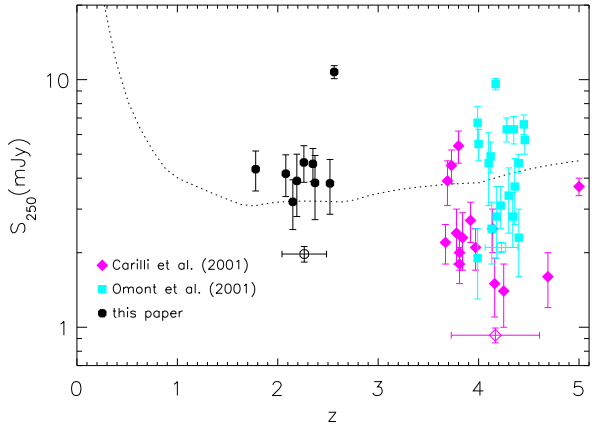},width=9cm}}
  \caption{ The redshift dependence of the 1.2~mm flux density ($S_{\rm 250}$) for
    the $z \approx 2$ quasars detected in this paper and the $z \geq
    4$ quasars reported by Omont et al. (2001) and Carilli et al.
    (2001a). The open symbols correspond to the mean redshift and
    weighted mean flux density of the respective total samples
    (detections and non-detections) - see Table 3.  The
    observational data are compared to the flux density expected from
    a $\rm 10^{13} \, L_\odot$ source with a spectral energy
    distribution similar to that of the starburst galaxy Arp~220.  }
  \label{figure2}
\end{figure}

With similar 1.2~mm flux densities, the implied luminosities, dust
masses and star formation rates for the $z \approx 2$ quasars are
comparable to the values derived for the SDSS and PSS samples.  This
is due to the fact that at high redshifts for a given FIR spectral
energy distribution (SED) and luminosity, the (sub)millimeter flux
density is nearly independent of redshift.  At comparable
sensitivities, there is no obvious difference in the 1.2~mm flux
densities of the $z \approx 2$ and $z \ga   4$ quasars (note though that
the $S_{\rm 250}$ flux densities correspond to rest-frame emission at 400
and $<$~240~$\rm \mu m$, respectively). This is illustrated in
Fig.~\ref{figure2} where the 1.2~mm flux densities of the $z \approx
2$ quasars and of the PSS and SDSS quasars are plotted against
redshift.  

%% Figure 2

                %% Figure 3
\begin{figure}[]
  \centerline{\psfig{figure={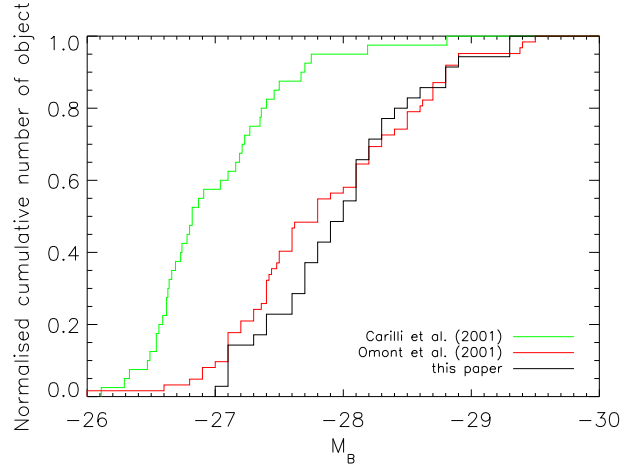},width=9cm}}
  \caption{The cumulative $\rm M_B$ distribution of the $z \approx 2$ QSO sample
    (this paper) compared to the distribution for the $z \geq 4$ QSO
    PSS and SDSS samples studied by Omont et al. (2001) and Carilli et
    al. (2001a), respectively.  }
  \label{figure3}
\end{figure}

                %% Table 3 
        %% Table 3 

\begin{table*}[tbhp]
\caption{Statistical properties of the 1.2~mm continuum surveys of the present QSO 
sample ($1.8 < z < 2.8$) and of the $z\geq4$ QSO PSS  (Omont et al. 2001) and SDSS 
(Carilli et al. 2001) samples  
         \label{statistics}} 
\begin{center}
\begin{tabular}{llrlcrrcr}
\hline \\[-0.2cm]
Sample &   & Numb. & Frac. & median r.m.s.& $\langle S_{250} \rangle$   & $\sigma_{S_{250}}$      & $\langle L_{FIR} \rangle$ & $ \langle M_B \rangle$ \\ 
  & & & (\%) & [mJy] & \multicolumn{2}{c}{[mJy]} & [$10^{12} \, L_\odot$] & \\
\hline \\[-0.2cm]

$z \approx 2$	& Detected   &   9   &  26   &  0.80 & 5.24 & 0.27  & 24.8 & $-$28.2 \\     
		& Undetec.   &  26   &  74   &  0.93 & 0.71 & 0.17  & 3.3 & $-$27.9 \\
		& Total      &  35   &       &  0.90 & 1.97 & 0.14  & 9.3 & $-$28.0 \\ 

PSS 		& Detected   &  19   &  29   &  0.70 & 4.52 & 0.16  & 15.7 & $-$27.8 \\
		& Undetec.   &  44   &  71   &  0.90 & 0.40 & 0.13  & 1.4 & $-$27.9 \\
		& Total	     &  63   &       &  0.80 & 2.07 & 0.10  & 7.2 & $-$27.9 \\

SDSS 		& Detected   &  15   &  38   &  0.40 & 2.35 & 0.11  & 8.3 & $-$27.0 \\
		& Undetec.   &  25   &  62   &  0.40 & 0.19 & 0.08  & 0.6 & $-$27.0 \\
		& Total	     &  40   &       &  0.40 & 0.93 & 0.06  & 3.2 & $-$27.0 \\ 

\hline \\[-0.2cm]
\end{tabular}
\end{center}
NOTE -- The average flux of the sample is $\langle S_{250} \rangle =
	\Sigma(w_iS_i)/\Sigma(w_i)$ where the weights are $w =
	1/\sigma^2$ and $\sigma_{S_{250}} = [\Sigma(w_i)]^{-0.5}$.
        Note that, in the cases of the $z\approx 2$ and PSS QSO samples,
        the average total flux densities and luminosities are biased 
	by the lower r.m.s. values of the detections.

\end{table*}

        \label{Table3}

The statistical properties of the 1.2~mm continuum surveys of the
present QSO sample ($1.8 < z < 2.8$) and of the PSS and SDSS samples
are compared in 
Table 3. Of the 35 $z \approx 2$ quasars
which were observed, 26\% were detected, which is comparable to the
detection fraction of 29\% of the PSS sample (both surveys have
comparable median r.m.s. values).  The 1.2~mm continuum survey of the
SDSS sample has about twice the depth of the $z \approx 2$ and PSS
samples, and a detection fraction of 38\%.
        
A parallel study of $z \approx 2$ quasars done at 850~$\rm \mu m$ by
Priddey et al. (2002) report a detection fraction of 15\%
for a median r.m.s. error of $\rm 2.8 \, mJy$, which is comparable to
the r.m.s. error of the present study if one adopts a typical ratio of
850~$\rm \mu m$ to 1.2~mm flux densities of 2.5.

Figure~\ref{figure3} shows the cumulative $\rm M_B$ distributions of the
observed sources for the $z \approx 2$ QSO sample compared to those for
the PSS and SDSS samples. It illustrates the difference in $\rm M_B$ range
of the SDSS sample, which has a $\rm M_B$ cumulative distribution fainter by
about one magnitude compared with that of the PSS and $z \approx 2$
samples (Fig.~\ref{figure3}). For the three samples, there is no obvious
difference in the cumulative magnitude distributions between the
detected sources and the whole sample (not shown in Fig.~\ref{figure3}).

\subsection{The Far-Infrared Luminosity Distributions}
\label{sec:statistics:luminosity}

The average 1.2~mm flux density, and hence $L_{\rm FIR}$, is smaller by
about a factor two for the SDSS sample compared with the PSS and $z \approx 2$
samples 
(Table 3 and Fig.~\ref{figure2}).  To statistically
explore the differences in the FIR luminosity distributions between
the samples, we applied the univariate methods of the survival
analysis (e.g., Isobe et al. 1986).  These statistical
tests compute the probability, $p$, that two samples derive from the
same parent distribution.

Comparing the PSS and SDSS samples, the tests return probabilities $p$
between 12 and 17\%, whereas for the PSS and $z \approx 2$ samples,
$p$ ranges between 25 and 38\%, and for the SDSS and $z \approx 2$
samples, from 1 to 5\%. These tests indicate that the PSS and $z
\approx 2$ samples could derive from similar parent FIR luminosity
distributions, whereas the SDSS sample could derive from a different
underlying distribution.  This difference is apparent already from its lower average flux densities 
(Table 3), and is even more
pronounced when considering the average luminosities.

That the millimeter fluxes measured for the SDSS QSOs are lower than
those found for the other samples could be related to the lower
optical luminosities of the SDSS QSO or to some other bias introduced
in the source selection. Some difference between the three samples
could also arise from the fact that the PSS and SDSS samples result
from homogeneous selections, whereas the $z \approx 2$ QSOs are
selected from the V\'eron-Cetty \& V\'eron catalogue, which is a
compilation of all QSOs discovered by various methods.

The FIR luminosities of the $z \approx 2$ QSOs are only slightly
higher than the luminosities of the $z \ga   4$ QSOs, especially when
comparing to the PSS sample which has a more similar optical
luminosity distribution, a fact which indicates that there is no
strong FIR luminosity evolution for optically luminous QSOs between
$2< z < 5$. This result differs from that obtained by Archibald et
al. (2001), who reported that the FIR luminosity of radio galaxies is
a strong function of redshift.

        %% Figure 4
        \begin{figure*}[ht!]
        \centerline{\psfig{figure={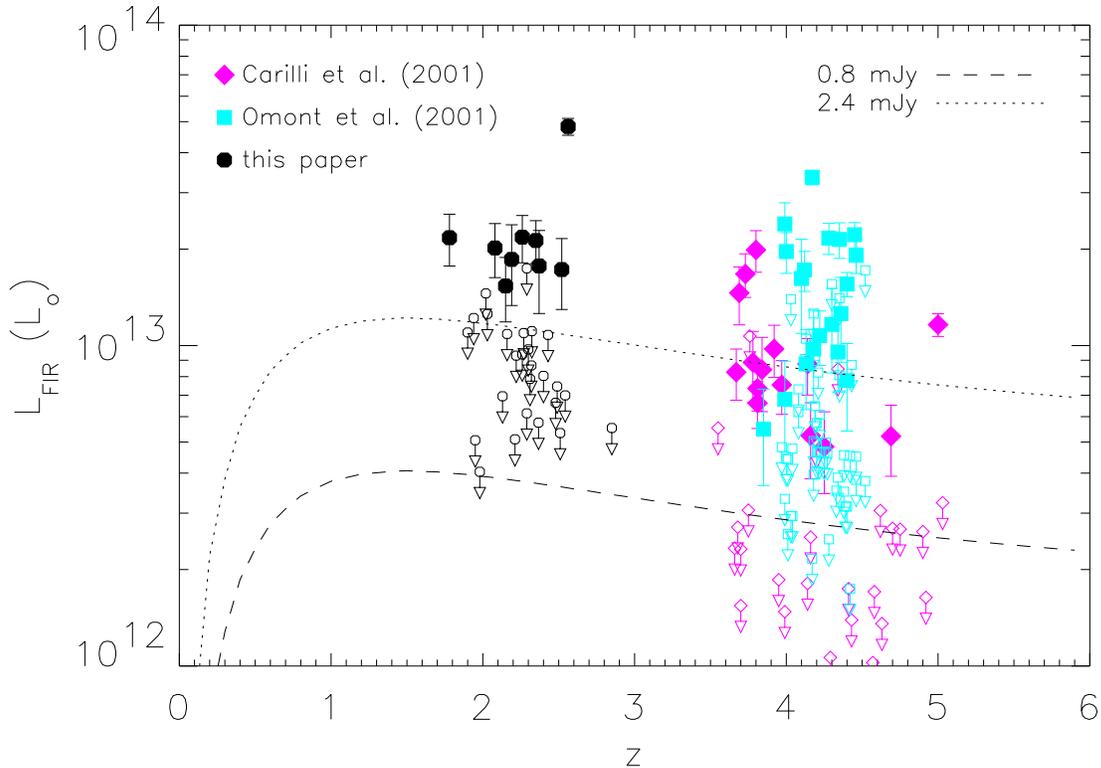},width=16cm}}
        \caption{Far-infrared luminosities, $L_{\rm FIR}$, implied by the
        MAMBO 1.2~mm (250~GHz) flux densities, plotted against the
        redshift $z$ for the 138 QSOs at $1.8 < z < 5.0$ reported by
        Carilli et al. (2001a), Omont et al. (2001) and this paper.
        The sources detected at 1.2~mm are shown as filled symbols,
        whereas the open symbols represent upper limits to $L_{\rm FIR}$
        (adopting the same definition as in Fig.~\ref{figure1}) for
        sources not detected at 1.2~mm.  The dashed line shows the
        typical r.m.s. flux error of the $z \approx 2$ and PSS 	  $z\geq4$ surveys
        (corresponding to $S_{\rm 250} = 0.8 \, \rm mJy$) and the dotted
        line shows the corresponding $3 \sigma$ detection limit.}
        \label{figure4}
        \end{figure*}

\subsection{Relation between the Far-Infrared and Optical Luminosities}
\label{sec:statistics:correlations}

We
applied the bivariate methods of the survival analysis to quantify the
probability of a correlation between the FIR and optical luminosities.  For the PSS and SDSS samples the tests
return high $p$ values indicating that there is no correlation between
$\rm M_B$ and $L_{\rm FIR}$. Specifically, the Cox regression, the generalized
Kendall's tau and the Spearman's rho\footnote{The small number of
sources, detected and undetected, is near the limit of the Spearman's
rho test accuracy} tests return values for $p$ of 55, 75 and 34\%,
respectively, for the PSS sample, and 77, 86, 94\%, respectively, for
the SDSS sample.  For the concatenated SDSS and PSS samples $p$ is
somewhat lower with 49, 41 and 15\%, respectively. Since $p$ is
significantly higher than 5\%, no correlation between
$\rm M_B$ and $L_{\rm FIR}$ can be proved.
The three bivariate methods applied to the $z \approx 2$ QSO sample
return $p=$ 22, 17 and 10\%, respectively, again showing no strong
correlation between $\rm M_B$ and $L_{\rm FIR}$.

The lack of an apparent correlation between the FIR and optical
luminosities agrees with previous studies (Omont et al.  2001; Carilli
et al. 2001a; Isaak et al. 2002; Bertoldi \& Cox 2002; Priddey et
al. 2002). However, the lower average values of $-M_B$ and $S_{\rm 250}$
for the the SDSS sample relative to the PSS and $z \approx 2$ samples is suggestive
that some correlation could be present, although the scatter of the present data
is too large to quantify this further. QSO samples with a wider range in $\rm M_B$ would
be required for further studies.  

Some proportionality between the FIR and the optical luminosities
might well be expected.  If the dust is heated by the AGN, both the FIR
and blue luminosities should scale with the bolometric luminosity, and
a strong correlation between the two should be seen, and is in fact
observed for low-redshift QSOs (Haas et al. 2000).
If on the other hand young stars are the main heating source of the
dust, a correlation between the {\it average} optical and FIR
luminosities could arise from the coeval growth of the central black
hole and of the stellar spheroid, which is also suggested by the
%%recent 
observation of a proportionality of the black hole and spheroid
masses in local galaxies (e.g., Kormendy \& Richstone 1995; Magorrian
et al. 1998).  A 
%%proportionality 
correlation between the optical and FIR
luminosities might be overshadowed by a strong scatter due to the
uncorrelated temporal variability of star formation and accretion
activity.

        %% Figure 5
        \begin{figure*}[ht!]
        \centerline{\psfig{figure={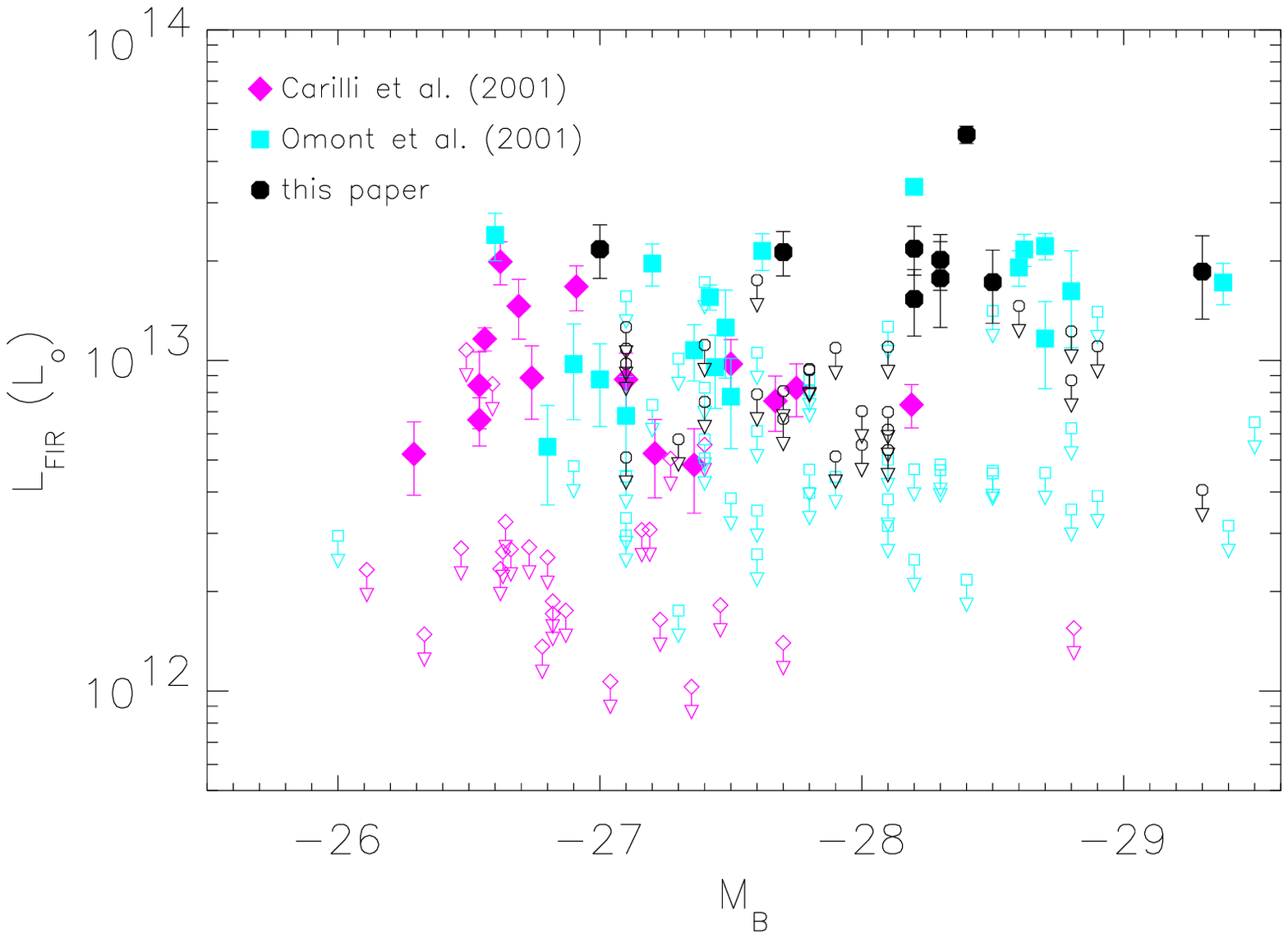},width=16cm}}
        \caption{ Far-infrared luminosities, $L_{\rm FIR}$, implied by the
          MAMBO 1.2~mm (250~GHz) flux densities as a function of
          rest-frame absolute B-band magnitude, $\rm M_B$, for the 138
          QSOs at $1.8 < z < 5.0$ reported by Carilli et al. (2001a),
          Omont et al. (2001) and in this paper (same symbols as in
          Fig.~\ref{figure4}).}
        \label{figure5}
        \end{figure*}

\section{Discussion}
\label{sec:discussion}

\subsection{Absence of Redshift Evolution of $L_{\rm FIR}$} 

The comparison of the 1.2~mm observations of QSOs at $z \approx 2$ and
$z \ga  4$ indicates that there is no strong evolution between $2< z < 5$ of the
FIR luminosity function of optically bright QSOs.  Quasars apparently differ
in this respect from radio galaxies which are FIR-brighter at higher
redshifts.
Archibald et al. (2001) argue that the apparent FIR luminosity
evolution of radio galaxies they found 
is not a selection effect related to the
inevitable youth 
of the known high-redshift radio galaxies.  Both their
luminosity and morphological evolution lead
to the interpretation that high-$z$ radio galaxies trace the formation and
evolution of massive elliptical galaxies (van Breugel et al. 1998;
Lacy et al. 2000; Pentericci et al. 2001).

It may be premature to claim a true difference between the QSOs and
high-$z$ radio galaxies. For one, we have yet observed only the most
optically luminous QSOs, which are not representative for the bulk of
the QSO emission at high redshift, which is due to QSOs with $\rm M_B\sim
-24,-25$.  Given the locally observed black hole to stellar bulge mass
ratios, the time-integrated rate of star formation (assuming that is
reflected by the FIR emission) vs. black hole growth (reflected by the
optical emission) must in fact be higher by a factor ten compared to
what we imply from the FIR to optical luminosity ratio of the QSOs in
our sample. This suggests that either the optically less luminous QSOs
show star formation rates similar to their bright peers, or the star
formation rate increases in time for the entire population -- a trend
which is not confirmed for our QSO sample, and contrary to what is
seen for the high-$z$ radio galaxies.

Another important unknown is the effective temperature of the
FIR emission -- which enters in high order when evaluating the FIR
luminosity, not even to mention a possible evolution of the dust emission
properties.

The lack of an evolution of the apparent FIR brightness (sampled at
rest wavelengths 400~$\mu$m and 240~$\mu$m at redshifts 2 and 4,
respectively) of optically bright QSOs might well be the cumulative
effect of several underlying, evolving processes, such as the decline
of star formation activity along with a drop in effective dust
%%AO
temperature, or opposite variations in the contributions of starbust and QSO radiation to dust heating.

Future studies should explore the FIR emission of optically fainter
QSOs, and try to measure the temperature of the FIR SED to better
estimate the FIR luminosity. To compare with an independent indicator
of star formation activity, the molecular line emission could be
measured. Currently, only the millimeter-brightest QSOs were observed,
and the few resulting CO line detections (e.g., Cox et al. 2002b) show no
redshift dependence of the line to continuum flux ratio.

\subsection{Source Counts and Relation between Far-Infrared and 
Optical Luminosities} \label{ref:counts}

From correlated SCUBA and Chandra X-ray observations it is estimated
that about $10-15\%$ of the submillimeter background sources contain
AGN (Bautz et al. 2000; Barger et al.  2001; Blain et al. 2002).  
%%AO
Most
of these sources are not optical quasars, but have
  optically obscured nuclei. 
Among $\sim$100 blank-field
sources detected with SCUBA and MAMBO, only two are known to be
optical quasars (Knudsen et al. 2001; Bertoldi et al. 2000a),

At millimeter wavelengths the density of background sources with
$S_{\rm 250} > 2$ mJy is observed to be $\rm \approx 1000 \, deg^{-2}$
(Bertoldi et al. 2000a; Blain et al. 2002).  The optically very bright
quasars we have studied with MAMBO should make a negligible
contribution to this millimeter background, as the following simple
estimate illustrates.  Fan et al.\ (2001a, b, c) infer a number
density $\rm \approx 2 \, deg^{-2}$ for QSOs with $\rm M_B < -26.5$ at all
redshifts, of which about one quarter would be brighter 2~mJy if we
assume that the millimeter brightness of QSOs is about constant from
$z=5-3.6$ (Carilli et al. 2001a; Omont et al. 2001) through $z \approx
2$ (this work) to the present. Optically very bright QSOs targeted in this study should thus only
contribute some 0.05\% to the millimeter background above 2 mJy.
However, there are $\sim 100$ times as many QSOs with fainter optical
luminosities $\rm M_B=-23$ to $-26.5$. If these had similar millimeter
properties as their brighter peers (a yet unfounded extrapolation),
such QSOs could contribute some 5 \% to the millimeter background, a
%%AO
number marginally consistent with the direct count of such objects from the SCUBA and MAMBO surveys and the analysis of the Chandra--millimeter associations. However, it seems that such counts favour a smaller contribution which would imply a slightly smaller QSO FIR  luminosity function in the range of $\rm M_B$$\approx -23$ to $-25$ than their brighter counterparts.

\section{Conclusion}
\label{sec:conclusion}

We have presented results of an on-going survey measuring the thermal
dust emission of high-redshift quasars. After surveying the millimeter
continuum of $z \ga   4$ QSOs (Omont et al. 2001; Carilli et al. 2001a),
we here report 250~GHz (1.2~mm) observations of optically luminous
($\rm M_B < -27.0$), radio-quiet quasars at redshift $1.8 < z < 2.8$,
corresponding to the peak of quasar activity. As for the $z \ga  4$
sources, dust masses of $\sim 10^8 \, M_\odot$ and FIR luminosities of
$L_{\rm FIR} \rm \sim 10^{13} \, L_\odot$ are implied by the observed
1.2~mm flux densities. If the FIR luminosity arises from massive star
formation, the inferred star formation rate would be $\approx 10^3 \,
\rm M_\odot \, yr^{-1}$.  The millimeter properties of the $z\approx
2$ quasars are similar to optically comparably bright QSOs at $z
\ga  4$. There is no sign of a redshift evolution of the FIR luminosity of
optically luminous QSOs, 
in contrast to what was reported for radio galaxies by Archibald et
al. (2001).

 With a significant sample size in the combined QSO samples at
redshifts $z \approx 2$ and $z \ga   4$, (43 detections and 95
non-detections), we find a large scatter and no clear correlation
between the millimeter and optical luminosities.  Lower average values
of $\rm M_B$ and S$_{\rm 250}$ for the SDSS sample compared with the PSS
sample are suggestive of a gross correlation, but the large scatter
makes this correlation of low statistical significance.

To better quantify the millimeter properties of high redshift QSOs and
especially the contribution of star formation to their emission, we
need to measure the millimeter emission of optically fainter QSOs. We further
need to  constrain the dust temperature through observations at
submillimeter and far-IR wavelengths, the AGN activity through observations at
far-IR, radio, and X-ray wavelengths, and we must measure the strength
of the molecular and atomic/ionic line emission. Most of these
observations will need to wait for a new generation of sensitive 
infrared/submillimeter observatories: SIRTF, Herschel and especially ALMA.

\medskip

\bigskip

{\it Acknowledgements:} We are most grateful to E. Kreysa and the
MPIfR bolometer group for providing MAMBO and to R. Zylka for creating
the MOPSI data reduction package.  Many thanks to the IRAM staff for
their support, and to all guest observers during the pool observing
sessions at the 30m.  J. Bergeron, C. De Breuck and P. Petitjean are
acknowledged for useful discussions, E.D. Feigelson for his help on
the survival analysis, and V. Junkkarinen for providing the analysis
of a Keck HIRES spectrum. The Referee's comments and suggestions by
C. Willott were very helpful to improve the manuscript.  This work was
carried out in the context of EARA, a European Association for
Research in Astronomy. The National Radio Astronomy Observatory (NRAO)
is a facility of the National Science Foundation, operated under
cooperative agreement by Associated Universities, Inc. IRAM is
supported by INSU/CNRS (France), MPG (Germany), and IGN (Spain).

\end{document}